\documentclass[aps,prd,showpacs,twocolumn,groupedaddress,sort&compress]{revtex4-1}

\usepackage{graphicx}
\usepackage{amsmath}

\newcommand{\Tr}{\mathrm{Tr}}

\begin{document}

\title{A covariant study of tensor mesons}


\author{A. Krassnigg}
\email[]{andreas.krassnigg@uni-graz.at}
\affiliation{Institut f\"ur Physik, Karl-Franzens-Universit\"at Graz, A-8010 Graz, Austria}

\author{M. Blank}
\email[]{martina.blank@uni-graz.at}
\affiliation{Institut f\"ur Physik, Karl-Franzens-Universit\"at Graz, A-8010 Graz, Austria}

\date{\today}

\begin{abstract}
We investigate tensor mesons as quark-antiquark bound states in a fully covariant Bethe-Salpeter equation.
As a first concrete step we report results for masses of $J^{PC}=2^{++}$ mesons from the chiral limit
up to bottomonium and sketch a comparison to experimental data.
All covariant structures of the fermion-antifermion system are taken into account and their roles and importance
discussed in two different bases. We also present
the general construction principle for covariant Bethe-Salpeter amplitudes of mesons with any spin and find
eight covariant structures for any $J>0$.
\end{abstract}


\maketitle

\section{Introduction\label{sec:intro}}


In QCD, mesons are viewed as bound states of (anti)quarks and gluons. Starting with a
$\bar{q}q$-picture they appear simpler than baryons and thus represent prime targets
for theoretical investigations. Spin and the corresponding meson degrees of freedom
are essential for an understanding of the meson spectrum and properties in general.

In a constituent-quark model
(e.\,g.~\cite{Godfrey:1985xj,Barnes:2005pb,Krassnigg:2003gh,Krassnigg:2004sp,Nefediev:2006bm,Lakhina:2006fy}),
mesons with total spin $J$ are easily obtained via adding units of
orbital angular momentum to a quark-antiquark state. In particular, given the quantum
numbers $J^{PC}$ for a meson with equal-mass constituents, the parity $P$ is given by $(-1)^{l+1}$
and the $C$-parity $C$ by $(-1)^{l+s}$. Furthermore, the total spin $J$, the internal (quark-antiquark) spin $s$,
and the orbital angular momentum $l$ and their projections have to satisfy the well-known
addition rules for angular momenta.

In the context of the Bethe-Salpeter equation (BSE), the Lorentz covariant structure of meson amplitudes
(also for arbitrary spin) has in the past been investigated mainly in setups involving reductions of the BSE
(e.g.~\cite{Smith:1969az,Aotsuka:1971tm,Fontannaz:1973jf,Mitra:1980rj,Nemirovsky:1983np,Qi:1986up,Jain:1993qh,Tiemeijer:1994bj,Olsson:1995zy,Koll:2000ke,Wagenbrunn:2007ie}).
Herein we present the first covariant study of tensor mesons that is consistent with respect to
the axial-vector Ward-Takahashi identity in the context of a Dyson-Schwinger--Bethe-Salpeter approach
to QCD.

The paper is organized as follows: Sec.~\ref{sec:bse} sketches the formalism used
and the corresponding details of immediate necessity,
Sec.~\ref{sec:amp} contains the explicit construction of the covariant amplitude for a $2^{++}$ meson, the construction
principle for $J>2$ amplitudes is given in Sec.~\ref{sec:higher}, the $2^{++}$ results are presented and discussed
in Sec.~\ref{sec:results}, and we conclude in Sec.~\ref{sec:co}.
All calculations have been performed in Euclidean momentum space.

\section{Mesons from the BSE\label{sec:bse}}

In this work, we employ QCD's Dyson-Schwinger-equations (DSEs) (see, e.g.~\cite{Fischer:2006ub,Roberts:2007jh}
for recent reviews) together with the quark-antiquark Bethe-Salpeter equation (BSE). The latter is
the covariant bound-state equation for the study of mesons in this context \cite{Smith:1969az}.
An analogous covariant approach to baryons is possible in a quark-diquark picture
(e.g.~\cite{Eichmann:2007nn,Nicmorus:2008vb,Eichmann:2010je} and references therein)
or a three-quark setup \cite{Eichmann:2009qa,Eichmann:2009zx}.

While the goal of a self-consistent solution of all DSEs can be held up in investigations of certain aspects of the theory
(see, e.\,g.~\cite{Alkofer:2008tt,Fischer:2008uz} and references therein), numerical hadron studies
such as ours require employment of a truncation. For our first covariant look at tensor mesons we use
the so-called rainbow-ladder (RL) truncation. It is both simple and offers the possibility for sophisticated
model studies of QCD within the DSE-BSE context, since it satisfies the relevant Ward-Takahashi identities (WTIs),
namely the axial-vector (see e.g.~\cite{Munczek:1994zz,Maris:1997hd}) and vector (see e.g.~\cite{Maris:1999bh,Maris:2000sk}) WTIs.
The literature regarding the employment of terms beyond RL truncation can be traced back from e.g.~\cite{Williams:2009ce,Chang:2010jq}.
The axial-vector WTI is essential to see chiral symmetry and its dynamical breaking correctly realized in the model
calculation from the very beginning. As the most prominent result, one satisfies Goldstone's theorem \cite{Munczek:1994zz}
and obtains a generalized Gell-Mann--Oakes--Renner relation valid
for all pseudoscalar mesons and all current-quark masses \cite{Maris:1997tm,Holl:2004fr}. We note that this relation
can be checked numerically and is satisfied at the per-mill level in our calculations.

The general structure of the BSE for a meson with spin $J$, total $q\bar{q}$ momentum $P$ and relative $q\bar{q}$
momentum $k$ or $q$, respectively, is
\begin{eqnarray}\label{eq:genbse}
\Gamma^{\mu\nu\ldots}(k;P)&=&\int^\Lambda_q\!\!\! K(k;q;P) S(q_+) \Gamma^{\mu\nu\ldots}(q;P) S(q_-)\,,
\end{eqnarray}
where the semicolon separates four-vector arguments. $\Gamma^{\mu\nu\ldots}(k;P)$ is the Bethe-Salpeter
amplitude (BSA) and has $J$ open Lorentz indices $\mu\nu\ldots$. The dressed-quark propagator $S(p)$ is
obtained from the quark DSE, the QCD gap equation. Since our focus here is the BSA, we refer the reader
to \cite{Maris:1997tm,Maris:1999nt,Krassnigg:2009zh} for more details on the quark DSE and to \cite{Krassnigg:2008gd} for
a description of our corresponding numerical solution method. In the BSE the quark and antiquark propagators
depend on the (anti)quark momenta $q_+ = q+\eta P$ and $q_- = q- (1-\eta) P$, where $\eta \in [0,1]$ is
a momentum partitioning parameter usually set to $1/2$ for systems of equal-mass constituents (which we do as well).
$\int^\Lambda_q=\int^\Lambda d^4q/(2\pi)^4$ represents a translationally invariant
regularization of the integral, with the regularization scale $\Lambda$ \cite{Maris:1997tm}.

The kernel $K$ in the homogeneous, ladder-truncated $q\bar{q}$ BSE is
essentially characterized by an effective interaction
$\mathcal{G}(s)$, $s:=(k-q)^2$. Following \cite{Krassnigg:2009zh}, an ansatz used
extensively for many years \cite{Maris:1999nt} is employed here, which reads
\begin{equation}\label{eq:interaction}
\frac{{\cal G}(s)}{s} = \frac{4\pi^2 D}{\omega^6} s\;\mathrm{e}^{-s/\omega^2}
+\frac{4\pi\;\gamma_m \pi\;\mathcal{F}(s) }{1/2 \ln [\tau\!+\!(1\!+\!s/\Lambda_\mathrm{QCD}^2)^2]}.
\end{equation}
This form provides the correct amount of dynamical chiral symmetry breaking as well as quark confinement
via the absence of a Lehmann representation for the dressed quark propagator. Furthermore, it produces the
correct perturbative limit, i.\,e.~it preserves the one-loop renormalization group behavior of QCD for
solutions of the quark DSE. As given in \cite{Maris:1999nt}, ${\cal F}(s)= [1 - \exp(-s/[4 m_t^2])]/s$, $m_t=0.5$~GeV,
$\tau={\rm e}^2-1$, $N_f=4$, $\Lambda_\mathrm{QCD}^{N_f=4}= 0.234\,{\rm GeV}$, and $\gamma_m=12/(33-2N_f)$.
Note that the same effective interaction appears also in the corresponding rainbow-truncated
quark DSE.

This function, which mimics the behavior of the product of quark-gluon vertex and gluon propagator, is
mainly phenomenologically motivated. While currently debated on principle grounds
(e.g.~\cite{Fischer:2008uz,Binosi:2009qm})
the impact of its particular form in the far IR on meson masses is expected to be small
(see also \cite{Blank:2010pa} for an exploratory study in this direction).

$D$ and $\omega$, in principle free parameters of the model interaction, can be used to investigate
certain aspects of both the interaction and the bound states in the BSE. In particular one
can interpret $D$ as an overall strength and $\omega$ as an inverse effective range of the interaction
(for more details and a thorough discussion of parameter dependence of the results see \cite{Krassnigg:2009zh}),
a notion first investigated in the study of radial meson excitations \cite{Holl:2004fr,Holl:2005vu}.
In the range $\omega\in [0.3,0.5]$ GeV, the prescription $D\times\omega=const.$ follows from
fitting of the model parameters to ground-state properties \cite{Maris:1999nt} and defines a one-parameter model,
which is the setup and range used in \cite{Krassnigg:2009zh} and also here. With all ingredients
specified, the BSE is solved numerically, a procedure well under control \cite{Blank:2010bp}.

\section{Tensor-meson Bethe-Salpeter amplitude\label{sec:amp}}

The BSA $\Gamma^{\mu\nu\ldots}(k;P)$ of a meson as a bound state of a quark-antiquark pair
depends on two four-vector variables: the total as well as the relative quark-antiquark
four-momenta $P$ and $q$, respectively. They can be parameterized in terms of the
Lorentz-invariant scalar products $P^2$, $q^2$, and $q\cdot P$.
The fermion-antifermion spin properties are encoded in the $4\times 4$ matrix structure of
$\Gamma^{\mu\nu\ldots}$ \cite{Smith:1969az}, where the open Lorentz indices appear in connection with the
total spin of the state. A corresponding basis of linearly independent structures $\{T_i^{\mu\nu\ldots}\}$ ($i=1,\ldots, N$)
involving Dirac matrices allows one to expand the BSA into a sum of Dirac covariants and the
corresponding scalar coefficients $F_i$, which
we will subsequently refer to as \emph{components} \cite{Blank:2010bp}. The latter only depend on the
aforementioned scalar products $P^2$, $q^2$, and $q\cdot P$, and one gets
\begin{eqnarray}
\Gamma^{\mu\nu\ldots}(k;P;\gamma)=\sum^N_{i=1}T_i^{\mu\nu\ldots}(k;P;\gamma)\; F_i(q^2,q\cdot P,P^2)\;,
\end{eqnarray}
where the dependence on $\gamma^\alpha$ has been made explicit and a generalized scalar product for the covariants
$T_i^{\mu\nu\ldots}$ is defined via the Dirac trace
\begin{equation}\label{eq:orthonormality}
\sum_{\mu\nu\ldots} \mathrm{Tr} [T_i^{\mu\nu\ldots} T_j^{\mu\nu\ldots}]=t_{ij}f(i,j)\;.
\end{equation}
One may also choose the basis elements orthogonal such that $t_{ij}=\delta_{ij}$, with the $f(i,j)$ functions
of $q^2$, $P^2$, and $q\cdot P$, or orthonormal such that in addition $f(i,j)=1$ for all $i,j$.
The sum is carried out over the $J$ indices $\mu,\nu,\ldots$.

Note that for an on-shell BSA $P^2=-M^2$ is fixed, while
one artificially varies $P^2$ in the solution process of the homogeneous BSE.
In the corresponding inhomogeneous BSE one has $P$ and therefore also $P^2$ as a completely
independent variable (see, e.g.~\cite{Bhagwat:2007rj,Blank:2010bp,Blank:2010sn}).

Thus, the on-shell scalar components $F_i(q^2,q\cdot P,P^2)$ effectively depend on the two variables
$q^2$ and $q\cdot P$, the latter of which can be parameterized by the variable $z\in [-1,1]$ related to
the cosine defining the angle between the four vectors $P$ and $q$. In principle, the components $F_i$
can be expanded further in Chebyshev polynomials, but we do not use such an expansion here (for
details and an illustration of Chebyshev moments, see \cite{Maris:1997tm,Krassnigg:2003dr}).

With the independent four momenta and $\gamma^\alpha$ one can construct four independent Lorentz-scalar structures,
\begin{equation}\label{eq:tsc}
\mathbf{1},\quad\gamma\cdot P,\quad\gamma\cdot q,\quad i\sigma^{q,P}\;,
\end{equation}
where $\sigma^{q,P}:=i/2\,[\gamma\cdot q,\gamma\cdot P]$. These four covariants, which
provide a basis corresponding to scalar mesons ($J^P=0^+$), serve as the basic
building blocks for any meson BSA. Together with pseudoscalar covariants ($J^P=0^-$) as well as
the bases for $J=1$ for all corresponding quantum numbers, these were explicitly constructed
in \cite{Krassnigg:2009zh}. Here we concentrate on $J=2$ and higher.

For $J=2$ one has 8 independent covariant structures in the BSA. Let
\begin{eqnarray}
q_\mu^T&:=&q_\mu-P_\mu \frac{q\cdot P}{P^2}\;,\\
\gamma_\mu^T&:=&\gamma_\mu-P_\mu \frac{\gamma\cdot P}{P^2}\;,\\
\gamma_\mu^{TT}&:=&\gamma_\mu-P_\mu \frac{\gamma\cdot P}{P^2} - q_\mu^T \frac{\gamma\cdot q^T}{(q^T)^2}\;,
\end{eqnarray}
be transverse projections of $\gamma$ and $q$ with respect to the total meson momentum $P$ and each other
(in particular the vectors $\{P_\mu\,,\;q_\mu^T\,,\;\gamma_\mu^{TT}\}$ are orthogonal to each other).

Defining furthermore the transverse projection of the metric
\begin{eqnarray}
g^T_{\mu\nu}=\delta_{\mu\nu}-\frac{P_\mu P_\nu}{P^2}
\end{eqnarray}
and the two transverse, symmetric, and traceless structures
\begin{eqnarray}\label{eq:m}
M_{\mu\nu}&=&\gamma_\mu^{T}q_\nu^T + q_\mu^T\gamma_\nu^{T}-\frac{2}{3}g^T_{\mu\nu}\;\gamma\cdot q^T\\\label{eq:n}
N_{\mu\nu}&=& \mathbf{1} ( q_\mu^T q_\nu^T-\frac{1}{3}g^T_{\mu\nu}\;q\cdot q^T )
\end{eqnarray}
one obtains the following set of tensor ($J^P=2^+$) covariants \cite{Smith:1969az}
\begin{multline}\label{eq:tensorcov}
T_1^{\mu\nu} = iM^{\mu\nu} \qquad T_2^{\mu\nu} = M^{\mu\nu}\;\gamma\cdot q\;q\cdot P-2N^{\mu\nu}\;q\cdot P\\
T_3^{\mu\nu} = M^{\mu\nu}\;\gamma\cdot P \qquad T_4^{\mu\nu} = 2 M^{\mu\nu}\,\sigma^{q,P}-4iN^{\mu\nu}\,\gamma\cdot P \\
T_5^{\mu\nu} = N^{\mu\nu} \qquad T_6^{\mu\nu} = iN^{\mu\nu}\;\gamma\cdot q \\
T_7^{\mu\nu} = i\,N^{\mu\nu}\;\gamma\cdot P\;q\cdot P \qquad T_8^{\mu\nu} = -2iN^{\mu\nu} \;\sigma^{q,P}
\end{multline}
Note that $T_5\ldots T_8$ were only given implicitly in \cite{Smith:1969az}.
All $T_i$ as given here are even under charge conjugation (for details, see e.g., \cite{Maris:1997tm,Krassnigg:2009zh}).
Thus, to obtain a $J^{PC}=2^{++}$ state, all components $F_i$ must be even functions of $q\cdot P$, which for
the present setup is indeed the property of the ground state in the system.
Note also that these covariants are in general neither orthogonal nor normalized;
orthonormal covariants can be generated via a Gram-Schmidt procedure applied to the set of terms in (\ref{eq:tsc}), leading to
\begin{equation}\label{eq:tscN}
\mathbf{1},\quad\gamma\cdot P,\quad\gamma\cdot q^T,\quad i\sigma^{q,P}\;.
\end{equation}
To orthogonalize the above $2^+$ covariants one introduces the symmetric and transverse expressions
\begin{eqnarray}
\tilde{M}_{\mu\nu}&=&\gamma_\mu^{TT}q_\nu^T + q_\mu^T\gamma_\nu^{TT}\;\;\textrm{and}\\
\tilde{N}_{\mu\nu}&=&q_\mu^T q_\nu^T\;,
\end{eqnarray}
which automatically satisfy Eq.~(\ref{eq:orthonormality}). The next step is to implement the
tracelessness, which is equivalent to orthogonality with respect to $g^T_{\mu\nu}$. This yields
\begin{eqnarray}\label{eq:m1}
M_{\mu\nu} = \tilde{M}_{\mu\nu}-g^T_{\mu\nu} \frac{\tilde{M}_{\rho\sigma}g^T_{\rho\sigma}}{(g^T)^2}\\\label{eq:n1}
N_{\mu\nu} = \tilde{N}_{\mu\nu}-g^T_{\mu\nu} \frac{\tilde{N}_{\rho\sigma}g^T_{\rho\sigma}}{(g^T)^2}\;,
\end{eqnarray}
which corresponds to Eqs.~(\ref{eq:m}) and (\ref{eq:n}),
and by multiplication with the four scalar covariants in (\ref{eq:tscN}) gives
the eight desired orthogonal tensor covariants. Note, however, that Eqs.~(\ref{eq:m}) and (\ref{eq:m1})
are slightly different.
Subsequently, normalization is achieved via $\hat{T}_i=T_i/\sqrt{\Tr[T_i \cdot T_i]}$.

\section{BSA for any meson spin\label{sec:higher}}
To consider mesons of any particular spin $J$, one has to construct Lorentz-tensors of rank $J$ which are
totally symmetric, transverse in all open indices and Lorentz-traceless (see, e.g., \cite{Corson:1982wi}):
such an object has the $2J+1$ spin degrees of freedom as demanded in quantum mechanics of a massive particle.
These restrictions, together with the properties of the Dirac matrices, lead to eight covariant structures
for $J\ge 1$. More precisely, the two tensors $M_{\mu\nu}$ and $N_{\mu\nu}$ defined above can be generalized
such that $N_{\mu\nu\ldots\tau}$ is the traceless part of
\begin{equation}\label{eq:generaln}
q_\mu^T q_\nu^T\;\ldots\;q_\tau^T
\end{equation}
and $M_{\mu\nu\ldots\tau}$ is the traceless part of the totally symmetric sum constructed from
\begin{equation}\label{eq:generalm}
\gamma_\mu^{TT} q_\nu^T\;\ldots\;q_\tau^T\;.
\end{equation}
Each of these multiplied by the four terms in (\ref{eq:tscN}) defines four rank-$J$ tensor covariants, in total
eight, orthogonal in the sense of Eq.~(\ref{eq:orthonormality}).

Obviously, Eqs.~(\ref{eq:m1}) and (\ref{eq:n1}) follow from this construction. As a further quick check
we consider the simplest such example, namely a vector meson: from $J=1$
one immediately obtains $N_\mu=q_\mu^T$ to give the first four, and
$M_\mu=\gamma_\mu^{TT}$ to give the second four covariants.

\section{Results\label{sec:results}}
Here we present results for $J^{PC}=2^{++}$ states that extend the study of Ref.~\cite{Krassnigg:2009zh}.
\begin{figure}
\includegraphics[height=\columnwidth,angle=270,clip=true]{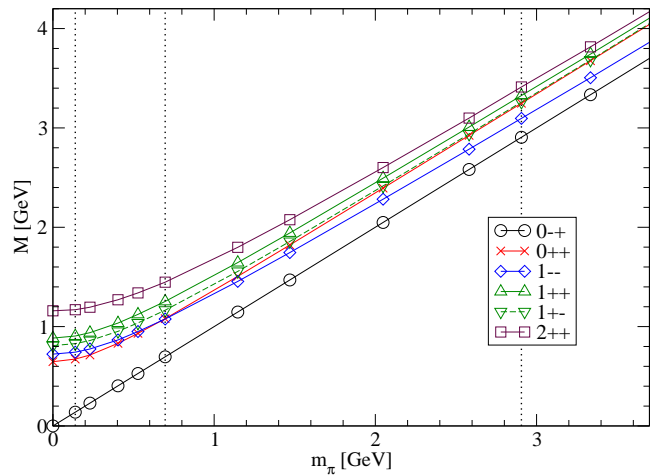}%
\caption{(Color online) dependence of meson masses on $m_\pi$ (the
pseudoscalar-meson mass calculated for a given current-quark mass). Vertical dotted lines
correspond to positions for light, strange, and charm $\bar{q}q$ states.\label{fig:massdep}}
\end{figure}
Consequently, we present correspondingly augmented figures here. Fig.~\ref{fig:massdep} shows the meson
masses for pseudoscalar, scalar, vector, axialvector, and tensor $q\bar{q}$ states as functions of the pion
mass, obtained from the BSE in RL truncation employing the effective interaction of Eq.~(\ref{eq:interaction}).
The three vertical dotted lines indicate the positions of the $n\bar{n}$, $s\bar{s}$, and $c\bar{c}$ states, respectively
(in the usual notation, $n$ here denotes light quarks).
Note that in RL truncation one cannot easily employ arbitrary flavor mixing between $SU(3)$-flavor octet
and singlet states; all states are either
purely $n\bar{n}$ or $s\bar{s}$, which corresponds to \emph{ideal} mixing.
As expected, the $2^{++}$ mass lies above all other states for the entire range from the chiral limit
to bottomonium.

\begin{figure}
\includegraphics[height=\columnwidth,angle=270,clip=true]{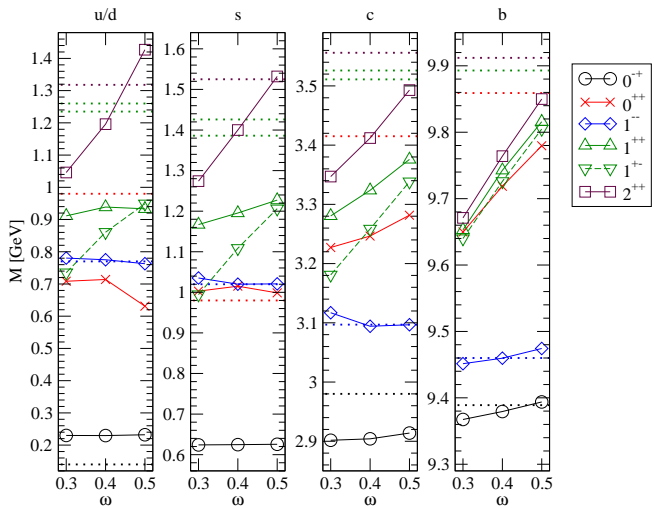}%
\caption{(Color online) dependence of meson masses on $\omega$. Dotted lines
correspond to experimental data \cite{Nakamura:2010zzi} where applicable. Figure updated
from Ref.~\cite{Krassnigg:2009zh}\label{fig:omegadep}}
\end{figure}

Comparison to experimental data is shown in Fig.~\ref{fig:omegadep} via the dotted lines in appropriate colors,
where for the $n\bar{n}$, $s\bar{s}$, $c\bar{c}$, and $b\bar{b}$ cases separately the dependence of the bound-state
masses on the model parameters is studied. More precisely, as mentioned in Sec.~\ref{sec:bse}, the inverse effective
range $\omega$ is used to explore the state's sensitivity to the details of the long-range part of
the strong interaction \cite{Holl:2004fr,Krassnigg:2009zh}. Two observations are noteworthy: First, the $2^{++}$
mass shows the same $\omega$-dependence as the other orbital excitations for each of the four columns.
Secondly, the agreement with experimental data is significantly better than for the often-quoted axial-vector states.
In the case of the latter, reconciliation of an RL study constrained by pseudoscalar- and vector-meson observables
seems unlikely, while this need apparently not be the case for tensor mesons, indicating that the latter are simpler
and not as sensitive to the details of the quark-gluon vertex employed in a DSE-BSE study as the axial-vector states.

A further technical note concerns the $2^{++}$ results for $\omega=0.5$ GeV: Due to the analytic structure
of the quark propagators for this parameter choice, the masses of the $2^{++}$ mesons are only accessible to
us via extrapolation techniques (see \cite{Bhagwat:2002tx} for a discussion). However, the extrapolations used
are reliable and stable, and the resulting error bars are smaller than the size of the symbols in Fig.~\ref{fig:omegadep}
except for the $u/d$ case, where we get an uncertainty of $\pm 75$ MeV.

An interesting question related to the issue of ``simplicity'' of the $2^{++}$ states in this approach is, how
important the various covariants/components are in the BSA or, in other words, how many covariants are needed to
arrive close to the full result. One possibility to investigate this is to leave out each covariant and
recompute the mass of the state with the remaining seven. Small differences to the full result then indicate
covariants of minor importance. Naturally there is a caveat for such an investigation, namely that the choice
of the covariants is somewhat arbitrary.

In our case we used two sets of covariants: the one given explicitly
above in Eq.~(\ref{eq:tensorcov}) and the other, orhonormal, constructed according to the principles detailed in Sec.~\ref{sec:higher}.
We have performed this test for both sets of covariants and present the results in Tab.~\ref{tab:covdiffs}.
We enumerate the orthonormal covariants in the following way: the four terms in (\ref{eq:tscN}) multiplied
with (\ref{eq:n1}) are numbered $1$ to $4$, and (\ref{eq:tscN}) multiplied with (\ref{eq:m1}) yield covariants
$5$ to $8$. For either set, one needs five of the eight covariants to arrive at a number which is within one percent of the
full result. Furthermore, omitting the contribution from $N^{\mu\nu}$ as indicated in \cite{Smith:1969az}
for this particular case yields a number which is $7$\% too low compared to the full result.

\begin{table*}
\caption{Meson mass of the $s\bar{s}$ $2^{++}$ state with $\omega=0.4$ GeV with all
covariants included as well as with single covariants left out. The change in bound-state
mass is given compared to the full result. The results are presented for
both the covariants of Eq.~(\ref{eq:tensorcov}) and the orthonormal set of covariants
constructed thereafter. All numbers are given in GeV.\label{tab:covdiffs}}
\begin{ruledtabular}
\begin{tabular}{rrrrrrrrrrr}
\multicolumn{2}{r}{Covariant missing} & none & 1 & 2 & 3 & 4 & 5 & 6 & 7 & 8 \\\hline
Eq.~(\ref{eq:tensorcov}) & Mass    & 1.448 & 1.575 & 1.455 & 1.502 & 1.509 & 1.502 & 1.287 & 1.452 & 1.450 \\
& Change  & +0.000 & +0.127 & +0.007 & +0.054 & +0.061 & +0.054 & -0.161 & +0.004 & +0.002 \\ \hline
Orthonormal& Mass    & 1.448 & 1.502 & 1.445 & 1.540 & 1.420 & 1.669 & 1.457 & 1.446 & 1.508 \\
& Change  & +0.000 & +0.054 & -0.003 & +0.092 & -0.028 & +0.221 & +0.009 & -0.002 & +0.060 \\
\end{tabular}
\end{ruledtabular}
\end{table*}

\section{Conclusions and outlook\label{sec:co}}

We have presented the complete set of Dirac covariants for mesons of spin $2$ and given an explicit construction
principle for the corresponding set of covariants for an arbitrary spin $J$ for the first time. We have
furthermore explored $2^{++}$ states in a well-established RL truncated model setup of QCD's DSEs and
solved the corresponding quark-antiquark BSE numerically. The results are both reasonable and surprising
in that they follow expected patterns, but are closer to experimental data than axialvector mesons,
even in the present simple setup.

The numerical calculation of further states with $J^{PC}=2^{-+},\,3^{--}$, etc.~is work in progress
and will be presented in future publications. Naturally, this includes radial excitations of these
states and opens up the concrete possibility to investigate Regge trajectories in the
covariant DSE-BSE approach.

\begin{acknowledgments}
We would like to acknowledge valuable discussions with R.~Alkofer, S.~Beni\'{c}, D.~Horvati\'{c}, and V.~Mader.
This work was supported by the Austrian Science Fund \emph{FWF} under project no.\ P20496-N16,  and
was performed in association with and supported in part by the \emph{FWF} doctoral program no.\ W1203-N08.
\end{acknowledgments}

%

\end{document}